# Ranking journals: Could Google Scholar Metrics be an alternative to Journal Citation Reports and Scimago Journal Rank?

**Emilio Delgado-López-Cózar & Álvaro Cabezas-Clavijo**

edelgado@ugr.es; acabezasclavijo@gmail.com

EC3: Evaluación de la Ciencia y de la Comunicación Científica, Departamento de Información y Comunicación, Universidad de Granada, Campus de Cartuja s/n E-18071 Granada (Spain)

**ABSTRACT**: The launch of Google Scholar Metrics as a tool for assessing scientific journals may be serious competition for Thomson Reuters' Journal Citation Reports, and for Scopus' powered Scimago Journal Rank. , A review of these bibliometric journal evaluation products is performed. We compare their main characteristics from different approaches: coverage, indexing policies, search and visualization, bibliometric indicators, results analysis options, economic cost and differences in their ranking of journals. Despite its shortcomings, Google Scholar Metrics is a helpful tool for authors and editors in identifying core journals. As an increasingly useful tool for ranking scientific journals, it may also challenge established journals products.

**Introduction**

If the launch of Google Scholar and Scopus in 2004 saw the end to the long-held monopoly of the Web of Science as the only scientific database offering bibliometric data, 2012 may well prove to have been another milestone with the launch and development of Google Scholar Metrics (GSM), a new product for evaluating the impact of scientific journals. This tool could be a serious competitor to the quintessential tool for evaluating journals, the Thomson Reuters' Journal Citation Reports (JCR) as well as to the more recent Scimago Journal Rank (SJR) developed by the Scimago Research Group and powered by Elsevier's Scopus database.

Such a blossoming of bibliographic tools has stimulated many studies analyzing the characteristics of Google Scholar as an information resource and as a tool for research evaluation[1-4], comparing it with the Web of Science and Scopus[5-14]. However, no study has been found comparing the characteristics of the bibliometric tools for rankings journals attached to such databases.

The present study compares these three products. Its aim is twofold. On the one hand, it analyzes the main characteristics and functionalities of each product. On the other, it compares the journal rankings offered by each of them. In this sense, the research question we address is quite clear: can Google position itself, with its new GSM, at the same level as Thomson Reuters and Elsevier?

**Overview of Google's Scholar Metrics, Journal Citation Reports and Scimago Journal Rank**

GSM was launched in April 2012 as a bibliometric tool, free of charge and to access, offering the H-Index for a wide range of scientific journals and other bibliographic sources (c. 40,000). The H-index is an extremely popular indicator amongst scientists but also shows many inconsistencies[15]. The most prominent one is that it is size-dependent, as its maximum value is limited by the total number of papers published by a journal. Obviously, it favours the most productive ones. Its main strength, besides ease of calculation (a journal has an h-index of n when n of its papers have at least n citations), are its stability, as it has no abrupt changes over time, and robustness, as it is very tolerant to errors in citations counting[16].

In its first edition, GSM offered the H-Index of papers published during 2007-2011, registering the citations they had received up to April 2012. The second edition updated the citation data to November 2012. The first analyses pointed out some methodological errors as well as some mistakes when processing the data[17-18] caused mainly by the semi-automatic methodology employed by Google to develop the product[19]. However, we must point out the rapidity and accuracy with which it retrieves data, the vast amount of sources it covers as well as the quick correction of many of these detected errors[20].

This tool aims at offering an alternative to the journal rankings in the market and particularly, to the JCR, the main yardstick for evaluating scientific journals and researchers by means of its Journal Impact Factor[21]. JCR is a bibliometric tool which offers various bibliometric indicators, - amongst them the impact factor – for more than 10000 journals which are, theoretically, the core of international scientific knowledge. Until 2007, when SJR[22] emerged based on data retrieved from Scopus, it was the only tool for such characteristics. The SJR is partly based on the Page Rank algorithm developed by Google. In fact, for this tool the reputation of the citing journal has a direct weight on the citation value. This is a significant difference from the impact factor, in which all citations count equally. Another basic difference is that, as data is derived from Scopus, which has a wider coverage, it allows calculating the impact indicator for a higher number of journals than those included in the JCR.

Next, we compare the main characteristics of these three products, concentrating on those features that one must take into account when selecting one or the other for bibliometric purposes, with a special focus on the recently launched GSM.

### *Coverage*

The coverage of a bibliometric tool (document types and total number of records indexed), as well as the indexing policies followed, are crucial when assessing the quality, exhaustiveness and precision of the indicators it offers. We find great differences between the three products under study. While GSM and SJR are multidisciplinary, JCR only offers the impact of journals from Science and Social Sciences, renouncing explicitly from its inception the calculation of the impact of the 1700 journals included in its Arts & Humanities Citation Index. Evidently, this means fewer journals covered (10677 in 2011) compared to SJR which includes 19708[23] journals, from which 2001 belong to the fields of the Arts & Humanities (10.2%). We do not know how many journals are covered by GSM as it is not formally declared. What we do know is that, according to its rankings by language, it calculates the H-Index of 5565 sources, of which 577 belong to the Arts & Humanities fields (10.4%). According to our calculations on the Spanish case[24-25], which accounts for 2.5% of the journals worldwide and has 901 journals indexed in GSM, we could estimate a total of c40,000 journals. As indicated elsewhere[26], this means that the GSM has double the number of journals indexed in the SJR and nearly triples the number offered by the JCR.

Another significant difference between GSM and the other traditional bibliometric tools is the document types it covers. While scientific journals constitute the main core of the rankings, it also includes conference proceedings, as well as collections and series from repositories (arXiv, SSRN, NBER, RePEC). Regarding the last, we should point out that GSM included in its first version the H-Index of various repositories such as arXiv or RePEC. This was considered a major error by Delgado López-Cózar & Robinson-García[18] and it was amended in its second edition. Instead of including whole repositories as single publications, they are now divided into the various series and collections in which they are structured. Therefore, in the case of arXiv, instead of showing the h-index of the repository one will find collections such as 'arXiv Astrophysics (astro-ph)' or 'arXiv Materials Science (cond-mat.mtrl-sci)' for instance, performing as individual units within GSM. JCR on the other hand, only includes scientific journals, while SJR also includes the impact of some conference proceedings in different research fields.

From a geographical perspective and in terms of languages other important differences exist between GSM and the other two products. We lack the precise data, but we know that GSM indexes journals from a wider range of places and languages than the others[27]. GSM presents rankings according to the 10 most representative languages of the world. This is because its data source, Google Scholar is less biased than the others to the English language[7]. However, though rankings of GSM can only be consulted for ten languages, it also calculates the H-Index of journals in other languages. Meanwhile, JCR and to a lesser extent SJR, are biased towards English-language journals as well as outlets from the United States and the United Kingdom, as can be judged by consulting Ulrich's Periodicals Index.

**Table 1: Coverage, indexing policies and citing sources**

| Characteristics | GSM | JCR | SJR |
|---|---|---|---|
| **Starting date** | 2012 | 1975 | 2007 |
| **Coverage** | | | |
| Areas | Multidisciplinary | Science & Technology; Social Sciences | Multidisciplinary |
| Arts & Humanities coverage | c. 4000 (estimation) (±10%) | 0 | 2001 journals (10.2%) |
| Type of documents | Journals, series and collections of repository documents, and Conference Proceedings | Journals | Journals and Conference Proceedings |
| # Sources | c. 40000 (estimation) 5565 are shown in rankings | 10677 | 19708 |
| # Countries | Unknown, but it is supposed to cover all countries | 82 | 98 |
| Anglo-Saxon bias (sources edited in US or UK) | Unknown | US: 4077; UK: 2048 (59.2%) | US: 5445; UK: 4914 (53.1%) |
| **Indexing policies** | | | |
| Criteria | No review. Publications with at least 100 articles published between 2007 and 2011 and that received any citations are indexed. | Review. Analysis of international publishing standards, thematic coverage, international scope, | Experts' review following these criteria: Journal Policy, Content, Journal Standing, Regularity |
| | | citation impact Acceptance rate c. 10-12%. | and Online Availability. Acceptance rate c. 40%. |
| Transparency about indexed sources | Lack of transparency. Indexed sources are not stated. | Transparency about indexed sources. Updated master lists are publicly available on its website. | Transparency about indexed sources. Updated master lists are publicly available on its website. |
| **Citing documents** | | | |
| Sources | Journals, proceedings papers, books, reports, dissertations, master's thesis, and any document with bibliographic references hosted in academic or publisher's websites, or repositories, which meet Google's inclusion criteria. | Web of Science-covered sources; mainly journals, but also proceedings and books. | Scopus-covered sources; mainly journals (only peer-reviewed material), but also proceedings and books. |
| Scientific nature | Its nature is academic but not strictly scientific. There are peer reviewed (journals, books, proceedings papers) and not peer reviewed sources (library guides, master theses, students' assignments, syllabuses, technical reports, teaching materials, presentations). | Its nature is mainly scientific but along with peer reviewed articles, not-peer reviewed materials (editorials and other type of items) from Web of Science covered sources can also be found. | Yes, only peer reviewed documents from Scopus covered sources are taken into account. |

*Indexing policies*

GSM has what could be described as a lax indexing policy and does not enforce any type of quality control over the journals it includes (see table 1). Its policy is in fact quite simple. It is limited to the quantity of papers published since it only calculates the H-Index of journals which have published at least 100 papers between 2007 and 2011 and excludes journals that have not received a single citation. Likewise, it indexes journal articles from the websites that follow Google's inclusion guidelines (mainly technical requirements), selected conference papers in Computer Science and Electrical Engineering and preprints from repositories such as arXiv, SSRN, NBER and RePEC[28].

On the other side we find the JCR, which has traditionally exercised a very demanding selection process, although it has relaxed such demands in recent years due to the expansion policy undertaken by the Web of Science in order to increase the coverage of non-English journals. Thomson Reuters states that only 10-12% of the approximately 2000 journals submitted for inclusion each year are finally indexed in the Web of Science (and hence, with the exception of the Arts & Humanities, in the JCR). The main aspects evaluated when considering the inclusion of a journal are: the fulfillment of the publication standards, scope of the journal and international representativeness, as well as citations received by the journal and especially, by its editorial board[29].

The expansion of the Web of Science database took place as a response to the larger coverage of Scopus. Scopus includes, as well as prestigious journals, others which have a limited quality and reach. Scopus states that it uses a selection process in which 'subject experts review titles using both quantitative and qualitative measures'. Criteria used by analysts are grouped into five main categories: Journal Policy, Content, Journal Standing, Regularity and Online Availability[30]. According to 2011 data, Scopus accepted 40% approximately of the journals which requested inclusion[31].

Both JCR and SJR are completely transparent regarding the sources they include. JCR allows the reader to browse and download all its journals for both Science and Social Sciences. SJR is the same i.e. it allows the reader to consult all the journals it indexes. This does not occur with GSM. As mentioned before, it lacks a master list, compromising its transparency, and this is thus one of its main weaknesses. It shows only the top 20 results of each query. This makes it practically impossible to know the whole set of sources for which the H-Index was calculated, and only a manual one-by-one query for all journals could establish that set. Such opacity is one of the main criticisms received not just by Google Scholar but also by its family of products[32-33].

### *Citing documents*

Another aspect in which significant differences can be found is related to the inclusion criteria of citing documents for the development of the rankings. JCR bases its calculations exclusively on journals, conference proceedings and monographs included in the Web of Science (but not only the peer-reviewed material of such sources). Scopus computes the citations directed from any document type included in its database and submitted to a peer review process (e.g., articles, reviews and conference papers). Once again, GSM follows a different path and lays down an open policy in which citations come from any document retrieved by Google Scholar (table 1). This means that the citation universe of GSM is bigger but also more uncontrolled than that of JCR and SJR, as on many occasions the nature of the citing documents is not strictly scientific. Therefore, as well as journals, books or proceedings, citations can also come from dissertations, master theses, students' assignments, syllabuses, technical reports, presentations or any other document type of an 'apparent' scientific content allocated within an academic website. According to a 2007 study, 12.8% of citations come from reports, thesis or sources other than books, conference papers and journal articles. Furthermore, 10.1% of mentions are from conferences, which may or may have not been peer-reviewed[34]. Such chaos can induce malpractices or fraudulent behaviours as it is easy to manipulate the citation data[35-36].

### *Bibliographic control and data standardisation*

Journals indexed in the JCR are classified into 232 subject categories according to the 2011 edition, with some journals classified in as many as six different categories. This classification is done intuitively, based upon visual examination of relevant citation data[37]. As for SJR, it is structured into 27 areas and 313 disciplines. Journals can be included in more than one discipline; however there are no indications on the maximum number of categories in which a journal can be indexed. Through random searches, we have detected some journals (such as Plos Computational Biology) that are included in up to seven subject areas, but we are unsure whether this is the maximum of disciplines for inclusion.. GSM classifies journals in 8 major areas and 313 subcategories, although these do not coincide with those in Scopus. Most of the journals are included in a single area and subcategory, although some journals from the field of Economics are included in two areas, while others, such as PLoS One or Nature Materials are in various subcategories. But due to the absence of a master list, it is not possible to know in how many areas or disciplines a journal can be. The data provided above is based on the top 20 journals displayed by GSM.

**Table 2: Bibliographic control and data standardisation**

| Characteristics | GSM | JCR | SJR |
|---|---|---|---|
| Data standardisation | Automatic information indexing. Journal titles are not standardised. | Standardisation of Journal titles, document types, address countries and language | Standardisation of Journal titles, document types, address countries and language |
| Subject classification | 313 disciplines | 232 disciplines | 313 disciplines |
| Does it state any criteria to classify journals? | No | 'Heuristic' method. Decision based upon visual examination of all relevant citation data. | No |
| Does it take any action against fraud or citation bias? | No, predatory journals can easily be found | Journals with anomalous citation patterns by an excessive concentration of citations are suppressed | SJR indicator does not take into account journals with self-citations above 33% |

Finally, another aspect of interest is whether these tools take any measures against fraud, especially concerning journals with anomalous citation patterns. JCR analyzes the self-citation rates of journals annually and suppresses those journals which inflate their rate, distorting their impact factor in order to position themselves in the journal rankings. The SJR adopts a different attitude and only takes into account self-citations when their rate is below a certain threshold (33%), avoiding possible inflation of the self-citation rate. GSM does not take any measures against these malpractices and in fact, it is possible to find in its rankings many so-called 'scam' journals according to Beall's list of predatory publishers.[38]

### *Search interface and visualisation of results*

On search interface and the visualisation of results, GSM is inspired by Google's usual simplicity, lacking many of the functionalities you will find in established bibliometric tools that include many and diverse options. The only exception in favour of GSM is the possibility of consulting the product in the reader's own language and not just in English - as you have to with the Thomson-Reuters and Scimago products. Other than that, these tools offer better features than Google's product.

**Table 3: Search interface and visualisation of results**

| Characteristics | GSM | JCR | SJR |
| --- | --- | --- | --- |
| **Search options** | Language, discipline (just for English and top 20 results) and journal title (all journals) | Journal (title, issn), subject, country, publisher | Journal (title, issn), subject (area & category), country, publisher |
| **Language** | All languages | English | English |
| **Journals rankings** | By Language, journal title words, area and discipline | By Journal, subject, country, publisher | By Journal, subject (area or category), and country |
| **Update frequency** | Not clear if it will be updated in a regular basis. First year, it has been updated in April and November. | Yearly | Yearly |
| **Number of results shown** | Top 100 publications per language<br>Top 20 publications per area or discipline (just for English publications)<br>Top 20 publications by title words | All journals. Every page shows 20 results | All journals. Every page shows 50 results |
| **Results ranking options** | The result is ranked by h-index. No other options can be found. | The results can be ranked by Journal Title, Total Cites, Impact Factor, 5-Year Impact Factor, total number of articles in the journal published in the JCR year, Immediacy Index, Half-life, Eigenfactor Score, Article Influence Score. | The results can be ranked by Journal Title, SJR, ), H index, Total Documents, Citable Documents (3 years), Cites per document (2 years and Total Cites (3 years). |
| **Data Download** | No option | Download data in txt format | Download data in MS Excel format |
| **Access to citing documents** | Just to papers which contribute to the journal H-index | No | No |
| **Access to citing journals** | Just for papers which contribute to the journal H-index | Yes (results are aggregated at journal-level) | No |
| **Access to cited journals** | No | Yes (results are aggregated at journal-level) | No |

GSM introduces a further novelty: rankings by publication language. However, this practice, not seen before in bibliometrics, is not practical as it generates other problems when indexing bilingual or multilingual journals[17]. GSM also offers the possibility to browse rankings by area and discipline (this option was included in the second edition of the GSM, November 2012), but only for the top 20 journals in English language. Furthermore, it offers the option of searching by source title. These features allow the reader to generate new rankings by area, discipline and title word (e.g. top 20 including the word 'publishing' in their title), which is another novel feature. JCR and SJR show bibliometric indicators of journals (searching for the title or the ISSN), disciplines, publishers or countries. In both these products, rankings are generated for each of the options (except for publishers in the SJR). Results are displayed in GSM depending on the options selected by the reader. Therefore, if the rankings are consulted by language, it offers a single page with 100 journals and, if consulted by Delgado-López-Cózar, Emilio and Cabezas-Clavijo, Álvaro (2013). Ranking journals: could Google Scholar Metrics be an alternative to Journal Citation Reports and Scimago Journal Rank? Learned Publishing, v. 26(2), 101-114. http://dx.doi.org/10.1087/20130206area or discipline, only the top 20 journals are shown. JCR shows the complete list of journals by ranking with 20 results by page. SJR displays results in the same way, but showing 50 by page instead of 20.

Another basic feature is the possibility of sorting the rankings by different criteria or indicators. In GSM results are presented ranked according to the H-Index and cannot be altered. JCR sorts by title, by any of the 8 bibliometric indicators displayed (amongst others the impact factor or the eigenfactor score) or by total citations. SJR allows sorting by title or by any of the six bibliometric indicators offered (amongst them the SJR indicator, the H-Index or citation average).

On GSM does not have annual editions of its rankings, as the other two products do, but a single time period (2007-2011 for the two editions launched in 2012). On the web, there is no information on whether the rankings will be updated on a regular basis. This seems unlikely given the strange dates on which the product was launched and updated, and the irregular citation windows (up to 1 April in the first edition and to 15 November in the second one). Another peculiarity of the product is that it does not maintain an archive with the journals' position and data of previous editions. Thus the update launched in November 2012 erased the previous version (April 2012), preventing any type of analysis of the historical evolution of journals.

The data in JCR are presented annually from 1975; the value of the impact factor (and the rest of the indicators) can vary each year. This data is published in June of each year and refers to the previous year. However, the data is revised establishing the final version free of errors in September. SJR follows a similar pattern and has published its rankings since 2007, going back to 1999 data; being updated twice each year (in April and September). An important issue is that 'with each data refresh, all values (current year and backwards) are recalculated and updated[39]. This is an important limitation of the indicator when establishing comparisons as the journals' values change with each new edition. On the other hand, SJR maintains an archive with the different updates made; therefore it is advisable, when establishing comparisons with this product, to indicate which edition is being referred to. In any case, both SJR and JCR show a great level of transparency maintaining an archive with the data of previous editions along with the possibility of downloading all data. GSM does not allow any type of download or export of its data.

The aspect in which GSM shows more transparency than the other tools is regarding the identification of citing documents, that is, the raw data on which the H-Index is based. Thus the reader can navigate from the H-Index of a journal to the citing documents that contribute to the indicator. This is not possible with JCR or SJR. These products offer the indicators but not the data which contributes to them. However, JCR, do offer all citing and cited journals, which is at least, a prior step to providing the citing document (table 3).

*Bibliometric Indicators*

On bibliometric indicators, GSM seems to have the motto 'less is more'. Contrary to the other products, GSM only offers two indicators: the H-Index, which is the one that ranks journals, and the H-Index Median Citations, in both cases for a five-year period. It is interesting that Google has made such a decision when other services offered by the company such as Google Scholar Citations offer other indicators such as total citations or the i-10 Index (number of papers with more than 10 citations). Moreover, it is a little disappointing that Google has not released its own metric indicator based for example on the algorithm used in the PageRank, as SJR does.

**Table 4: Bibliometric indicators**

| Characteristics | GSM | JCR | SJR |
|---|---|---|---|
| **Main bibliometric Indicator** | H index | Impact Factor | SJR |
| **Period for calculation** | 5 years | 2 years | 3 years |
| **Does it take into account self-cites for calculation?** | Yes | Yes, but it is also possible to know the impact factor without self-citations | Yes, up to a third of the total share; it discounts any above this threshold |
| **Other bibliometric Indicators** | Median h index | 5-Year Impact Factor, Total Cites, Citable items, Immediacy Index, Half-life, Eigenfactor Score, Article Influence Score | H index, Total Documents (current year), Total Documents (3 years), Citable Documents (3 years), Cites per document (2 years), Total Cites (3 years), Average number of references per document |
| **Results analysis for journals** | No | Position within subject categories, Impact Factor Trend Graph, Cited Journal Graph, Cited Journal Data, Citing Journal Graph, Citing Journal Data, Journal Self Cites, Related journals, Source Data: number articles, reviews, other items; Number of references, ratio references per articles | Trend Graph SJR indicator vs. Cites per Doc (2y), Citation vs. Self-Citation, Cites per Document vs. External Cites per Document, Cites per Document in 2, 3 and 4 years windows, Journal's Citable vs. Non Citable Documents, International Collaboration, Journal's Cited vs. Uncited Documents, Journal's Citable vs. Non Citable Documents |
| **Journals comparison** | No | No | Yes, 4 journals' metrics can be analysed together |
| **Bibliometric Indicators for subject categories** | No | Total Journals in category, Total Articles, Total Cites, Median Impact Factor, Aggregate Impact Factor, Aggregate Immediacy Index, Aggregate Cited Half-Life | No |
| **Results analysis for subject categories** | No | Impact Factor Trend Graph, Cited Journal Graph, Cited Journal Data, Citing Journal Graph, Citing Journal Data, Journal Self Cites, Related journals, Source Data: number articles, reviews, other items; Number of references, ratio references per articles | No |
| **Explanation of indicator's calculation** | Yes | Yes | Yes |

Such a basic bibliometric toolkit displayed by GSM stands in contrast to JCR and SJR, which offer up to 8 different indicators per journal, always emphasizing one of them, - impact factor and SJR respectively. As well as showing other indicators, it is important that they also offer other bibliometric services or tools offering added-value (see table 4). GSM lacks of any of these tools other than showing the citing documents that contribute to the H-Index of each journal. JCR offers a complete profile of each journal and year, enumerating among other aspects the bibliographic data (ISSN, Publisher, subject areas), the journal's quartile for each area, the impact factor trend, related journals and the percentage of self-citations. These options are also available at a subject-category-level. SJR also offers a lot of information for each journal, giving the number of citations and references, percentage of self-citations, international collaboration, etc. Also, it allows comparisons between up to four journals, an option which is unavailable in JCR. Certainly these features would only interest bibliometricians and subject-experts, but creating products such as these implies offering capabilities for professional use.

All three products include help pages which explain how each indicator was calculated. GSM has less detail. The other two also have different brochures and scientific papers in which further details about the technicalities of the indicators are given.

**Cost**

Ability to access and cost are of course significant factors. GSM, following Google's usual policy, is free and easy to access. JCR is not free. Both the Web of Science databases as well as JCR are closed products to which access is available on subscription. This cost varies depending on the institution and the products enrolled. For instance, Spain paid 4 million Dollars in 2012[40] for full access to the Web of Science (including JCR) for all the public universities and research institutions in the country. SJR and Scopus sit between these two extremes. Although SJR is free to access, Scopus is only available through subscription.

**Table 5: Cost**

| Characteristics | GSM | JCR | SJR |
|---|---|---|---|
| **Access and Cost of Journal product and data source** | Open-access product for both; the journal rankings and Google Scholar | Access to JCR and Web of Science data is by paid-license. Costs vary | The journal rankings are Open-access but not the Scopus database. Costs vary |

**Journal rankings: results' comparison**

Clearly a main issue is whether GSM's new indicator-based rankings are similar to or different from those that already exist, and to what extent they measure something different. These tests can be performed in multiple ways[41-42] but basically they aim at comparing journals' datasets, and examining the extent of correlation.

Previously correlations between JCR and SJR have been studied[43-46], and now the new Google product is introduced. We have downloaded (date: 7-15 January 2013) all journals shown in GSM rankings for the 2007-2011 period (N=5567), all JCR journals in the 2011 edition (N=10667) and all SJR journals for the same year (N=19708).

Titles from the three datasets have been examined. Through a matching process (which was rather difficult as GSM doesn't show journals' ISSN) we have been able to identify the 3423 journals that are included in all these populations. We applied the Spearman correlation coefficient (rho). This is a statistical measure that is often used to calculate the degree of association between two variables[47]. The statistical analysis of data for all the indicators was carried out with SPSS v 20.0.0.

Results show a very strong correlation between the two GSM indicators, which leads us to think that the indicator of 'median of citations' does not give a different bibliometric vision of journal rankings. The inclusion of such an indicator can only be justified as a way to sort those journals with the same h-index. Regarding correlations with the main indicators of each product, both JCR's impact factor and SJR are highly correlated with GSM h-index, being 0.818 in the first case and 0.778, in the second case (table 6). It also calls to our attention that there are no significant differences between the traditional impact factor (which uses a 2-year citation window) and the 5-years impact factor (0.82) when comparing to GSM's h-index.

**Table 6: Spearman correlation between GSM and JCR/SJR indicators**

| PRODUCT | INDICATOR | H_INDEX | MEDIAN_CITES |
|---|---|---|---|
| GSM | H Index | 1,000 | 0,975 |
| | Median Cites | **0,975** | 1,000 |
| JCR | Total Cites | **0,867** | **0,812** |
| | Impact Factor | **0,818** | **0,803** |
| | 5years Impact Factor | **0,82** | **0,813** |
| | Jcr Inmediacy | 0,679 | 0,665 |
| | Total Articles | 0,633 | 0,545 |
| | Cited Half Life | 0,105 | 0,094 |
| | Eigenfactor | **0,905** | **0,856** |
| | Article Influence | 0,755 | 0,77 |
| SJR | Sjr | 0,778 | 0,795 |
| | H Index | **0,877** | **0,84** |
| | Total Docs 2011 | 0,638 | 0,553 |
| | Total Docs 3 Years | 0,664 | 0,576 |
| | Total References | 0,736 | 0,664 |
| | Total Cites 3 Years | **0,881** | **0,819** |
| | Citable Docs 3 Years | 0,657 | 0,567 |
| | Cites Doc 2 Years | **0,834** | **0,825** |
| | References Per Doc | 0,134 | 0,164 |

N= 3423 journals. All correlations are significant at the level of $p < 0.01$. In bold, correlations >0.8

Likewise, there is also a high correlation between GSM's and SJR's h-index (0.877), despite of the significant difference in the citation window (5 years for GSM and 13 for SJR). These data support the high similarity between rankings, regardless which data source is used and is consistent with findings from different analyses for samples from journals in Critical Care, Communication, Library and Information Science, Forestry or Economics & Business[16,26,27,48-49] amongst others.

Furthermore, a high correlation has also been found between GSM's h-index and size-dependent indicators, such as JCR total cites (0.867) or the eigenfactor indicator (0.905), something that is not surprising, as the h-index is a measure that is strongly affected by the size output too.

In summary, what these data is telling us is that despite of the lack of control and standardisation, GSM ranks journals in a very similar way to JCR and SJR and therefore, generally speaking, it is a reliable and valid alternative to traditional indexes when measuring journal's impact. Of course, this does not mean that they are identical; journals with an unusually high number of papers will be positioned in much higher positions in GSM that in the other products, whereas low output journals will be benefited by relative indicators such as impact factor or SJR.

For instance PNAS, which is ranked sixth according to GSM drops down to position 131 in JCR, whereas PLoS One, which is positioned as number 52 in GSM tumbles to position 800 when using the JCR. As Leydesdorff[42] states, the h-index leads to counter-intuitive results 'because of the attempt to bring the size component and the impact component under a single denominator'. Important distortions are also due to other reasons, such as the citation window. Thus American Economic Review which is ranked 44th according to GSM (and 69th in SJR) falls down to position 1168 in JCR, as a result of the short citation window used by this tool (2 years). In conclusion; rankings are similar but not the same.

**Conclusions**

We have analyzed the three main journal evaluation tools from different approaches. This study is justified by the launch of a new player, GSM, by Google, which may be a serious competitor for the traditional reference, Thomson Reuters' JCR and for the more recent Scopus' powered SJR. Six points deserve to be highlighted from this work.

- GSM is a different tool, both in its conception, - as it is a hybrid product

(bibliometric on the one hand and bibliographic on the other), not as JCR and SJR, which are purely bibliometric, - and in its implementation (indexing policies, coverage, architecture and formal presentation).

- GSM is simple and easy to use and understand by any scholar, and uses indicators which can be effortlessly calculated and replicated, contrary to JCR and SJR which are complex, hard to use and which require an expert knowledge to fully interpret them. Nevertheless, this product is influenced by Google's philosophy of simplicity as confirmed by the indicator chosen; the h index, which is hugely popular for scientists due to its simplicity and its (apparently) good performance in ranking scientists. It is unquestionable that Google's intrusion has contributed to the popularisation of Bibliometrics[16], as it is now accessible to everyone within academia. In just a few minutes, any scholar can find papers relevant to their research topic, learn a journal's impact or even set up their scientific profile and know instantly which their top-cited papers are. The same happens with journals' editors, who can get for free objective information about their performance (and competitors' performance). GSM also informs the reader about top-cited papers in the discipline, outstanding authors or topics of interest. JCR and SJR, as well as the databases in which they are based (Web of Science and Scopus) cannot compete with GSM in this aspect. However, GSM does not meet the expectations of bibliometricians.

- GSM is a product which lacks transparency; it does not incorporate any scientific control regarding their selection policy or data processing, in contrast to the control exerted by JCR and SJR. This collides with the minimum requirements demanded of any scientific tool, especially by the bibliometric community which needs transparent tools when building any valid bibliometric indicator. But, given the rapid evolution of GSM[20] these shortcomings may be solved in the near future. It is not clear that the scientific community, which requires previous control of certified knowledge (as peer review shows), will be inclined to accept a product which stands out by its lack of regulation.

- GSM does not take any action against potential data manipulation, especially of those concerning citations[36]. This is a crucial problem and could invalidate GSM

- as a source for assessing journals. It would be desirable to implement some kind of control against fraud which would prevent authors or editors from malpractice.
- Despite the aforementioned limitations, GSM offers very similar rankings to those of traditional data sources. Core journals can not only be identified with GSM, but its dataset is much more representative of the world's research activity, contrary to the portrait displayed by JCR and SJR. Consequently, and regardless of the indiscriminate coverage of academic material, GSM's results appear to be highly reliable.
- GSM is a free product, which represents an important difference when comparing it with JCR and the Web of Science business model or the hybrid model of SJR and Scopus.

In conclusion, if the results derived from GSM are very similar to those which can be obtained from paid sources, many institutions may consider using only Google's product. If GSM can sort out its many shortcomings, and consolidate itself as a useful tool for authors and editors, there will be a real challenge to established journal ranking products.

**Acknowledgments**

This work has been funded under the project HAR2011- 30383-C02-02 of the Dirección General de Investigación y Gestión of the Plan Nacional de I+D+I. Ministerio de Economía y Competitividad (Spain).
The authors would like to thank Nicolás Robinson-García for translating this text and Alberto Martín-Martín for journals data processing